\documentclass[letterpaper,conference]{IEEEtran}
\IEEEoverridecommandlockouts
\usepackage{graphicx}
\usepackage[utf8]{inputenc} 
\usepackage[T1]{fontenc}
\usepackage{url}
\usepackage{ifthen}
\usepackage{cite}
\usepackage{color}

\usepackage{cite}
\ifCLASSINFOpdf
\else
\fi
\usepackage{array}
\usepackage{graphicx}
\usepackage{hyperref}
\usepackage{amsmath}
\usepackage{amssymb}
\usepackage{bm}
\usepackage{bbm}
\usepackage{algorithmicx}
\usepackage{algorithm}
\usepackage{algpseudocode}
\usepackage{array}
\usepackage{booktabs}
\usepackage{multirow}
\usepackage{makecell}
\usepackage{tikz}
\usetikzlibrary{shapes.geometric, arrows}

\begin{document}
%
% paper title
% Titles are generally capitalized except for words such as a, an, and, as,
% at, but, by, for, in, nor, of, on, or, the, to and up, which are usually
% not capitalized unless they are the first or last word of the title.
% Linebreaks \\ can be used within to get better formatting as desired.
% Do not put math or special symbols in the title.
% \title{List Viterbi with Code-Specific CRC: Managing Undetected Errors, NACKs, and Complexity}
\title{CRC-Aided List Decoding of Convolutional and Polar Codes for Short Messages in 5G}

% author names and affiliations
% use a multiple column layout for up to three different
% affiliations
%author{\IEEEauthorblockN{Jacob King, William Ryan, and Richard~D.~Wesel}
%\IEEEauthorblockA{ Department of Electrical and Computer Engineering\\
%University of California, Los Angeles, Los Angeles, CA 90095, USA\\
%Email: \{emliang, hengjie.yang, wesel\}@ucla.edu}
%}

% conference papers do not typically use \thanks and this command
% is locked out in conference mode. If really needed, such as for
% the acknowledgment of grants, issue a \IEEEoverridecommandlockouts
% after \documentclass

% for over three affiliations, or if they all won't fit within the width
% of the page, use this alternative format:
% 

\author{\IEEEauthorblockN{Jacob King\IEEEauthorrefmark{1},
Alexandra Kwon\IEEEauthorrefmark{1},
Hengjie Yang\IEEEauthorrefmark{1},
William Ryan\IEEEauthorrefmark{2}, and
Richard D. Wesel\IEEEauthorrefmark{1}
}
\IEEEauthorblockA{\IEEEauthorrefmark{1}Department of Electrical and Computer Engineering, University of California, Los Angeles, Los Angeles, CA 90095, USA}
\IEEEauthorblockA{\IEEEauthorrefmark{2}Zeta Associates, Aurora, Denver, CO 80011, USA}
Email: jacob.king@ucla.edu, alexandrakwon@ucla.edu, hengjie.yang@ucla.edu, ryan-william@zai.com, wesel@ucla.edu,}

% use for special paper notices
%\IEEEspecialpapernotice{(Invited Paper)}

% make the title area
\maketitle

% As a general rule, do not put math, special symbols or citations
% in the abstract
\begin{abstract}
This paper explores list decoding of convolutional and polar codes for short messages such as those found in the 5G physical broadcast channel.  A cyclic redundancy check (CRC) is used to select a codeword from a list of likely codewords. One example in the 5G standard encodes a 32-bit message with a 24-bit CRC and a 512-bit polar code with additional bits added by repetition to achieve a very low rate of 32/864.  This paper shows that optimizing the CRC length improves the $E_b/N_0$ performance of this polar code, where $E_b/N_0$ is the ratio of the energy per data bit to the noise power spectral density.  Furthermore, even better $E_b/N_0$ performance is achieved by replacing the polar code with a tail-biting convolutional code (TBCC) with a distance-spectrum-optimal (DSO) CRC.  This paper identifies the optimal CRC length to minimize the frame error rate (FER) of a rate-1/5 TBCC at a specific value of $E_b/N_0$.  We also show that this optimized TBCC/CRC can attain the same excellent $E_b/N_0$ performance with the very low rate of 32/864 of the 5G polar code, where the low rate is achieved through repetition. We show that the proposed TBCC/CRC concatenated code outperforms the PBCH polar code described in the 5G standard both in terms of FER and decoding run time. We also explore the tradeoff between undetected error rate and erasure rate as the CRC size varies. 

\end{abstract}

%{\let\thefootnote\relax\footnote{{This research is supported in part by National Science Foundation (NSF) grant CCF-1618272 and a grant from the Physical Optics Corporation (POC). Any opinions, findings, and conclusions or recommendations expressed in this material are those of the author(s) and do not necessarily reflect the views of the NSF or POC. Research was carried out in part at the Jet Propulsion Laboratory, California Institute of Technology, under a contract with NASA.}}}

% For peer review papers, you can put extra information on the cover
% page as needed:
% \ifCLASSOPTIONpeerreview
% \begin{center} \bfseries EDICS Category: 3-BBND \end{center}
% \fi
%
% For peerreview papers, this IEEEtran command inserts a page break and
% creates the second title. It will be ignored for other modes.
\IEEEpeerreviewmaketitle
{\let\thefootnote\relax\footnote{{This research is supported by Zeta Associates Inc. and National Science Foundation (NSF) grant CCF-2008918. Any opinions, findings, and conclusions or recommendations expressed in this material are those of the author(s) and do not necessarily reflect views of Zeta Associates Inc. or NSF.}}}
\section{Introduction}
% no \IEEEPARstart
% You must have at least 2 lines in the paragraph with the drop letter
% (should never be an issue)

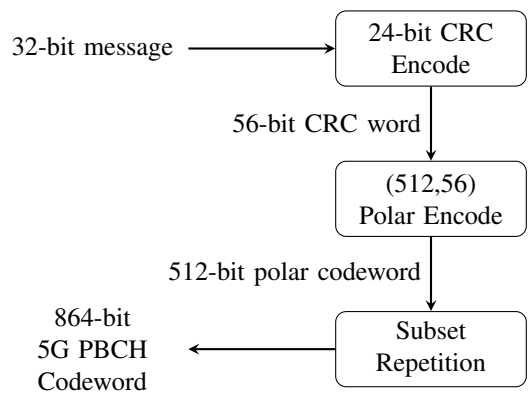
\begin{figure}

    \centering
    
    \tikzstyle{box} = [rectangle, rounded corners, minimum width=2cm, minimum height=1cm,text centered, text width=2.3cm, draw=black]
    \tikzstyle{no box} = [text centered, text width=2.3cm]
    \tikzstyle{arrow} = [thick,->,>=stealth]
    
    \begin{tikzpicture}[node distance=2cm]
    %\node (box1) [box] {32-bit message};
    \node (box1) [no box] {32-bit message};
    
    \node (box2) [box, right of=box1, xshift=2.5cm] {24-bit CRC Encode};
    \node (box3) [box, below of=box2] {(512,56) Polar Encode};
    \node (box4) [box, below of=box3] {Subset Repetition};
    %\node (box5) [box, left of=box4, xshift=-2.5cm] {864-bit 5G PBCH Codeword};
    \node (box5) [no box, left of=box4, xshift=-2.5cm] {864-bit 5G PBCH Codeword};
    
    \draw [arrow] (box1) -- (box2);
    \draw [arrow] (box2) -- node [anchor=east] {56-bit CRC word} (box3);
    \draw [arrow] (box3) -- node [anchor=east] {512-bit polar codeword} (box4);
    \draw [arrow] (box4) -- (box5);
    \end{tikzpicture}

    \caption{Block diagram of 5G PBCH polar encoding scheme.  The PBCH uses a 24 bit CRC, then polar encodes to 512 bits before applying repetition to get to 864 bits.}
    \label{fig:5GPBCH}
\end{figure}

Polar codes have seen wide interest since Arikan first described the paradigm and showed that it could achieve channel capacity \cite{ArikanPolar}.  Polar codes have found application in the physical broadcast channel (PBCH) of the 5G standard \cite{3GPP38.212}, \cite{Polar5G}.  In particular, Fig. \ref{fig:5GPBCH} shows how a polar code is used to transmit a 32-bit message over the 5G PBCH.  First the 32-bit message is protected by a 24-bit CRC to provide a 56-bit input to the polar code.  The polar code produces 512 bits, which are augmented by repetition to produce an 864-bit 5G PBCH codeword. This paper explores ways to improve the frame error rate (FER) vs. $E_b/N_0$ performance of this 5G PBCH code and considers alternatives.  

For example, FER vs $E_b/N_0$ improvement is achieved by reducing the length of the CRC.  CRCs, as described in \cite{Peterson1961}, are very powerful as error detecting outer codes.  However, for the best FER vs $E_b/N_0$ performance, the error detection benefit provided by the CRC  needs to be balanced with the corresponding overhead requirement.  We show that replacing the 24-bit CRC with the smaller 11-bit or 12-bit CRC increases the number of frozen bits and reduces the FER for a given $E_b/N_0$.

%Generally, an error correcting code with a lower code rate will have a lower error rate than one with a higher code rate.  This makes it difficult to compare codes of different rates, as the lower rate codes will have an advantage over the higher rate codes.  However, $E_b/N_0$ adjusts for code rate, so lower rate codes will have a higher noise level than higher rate codes for equivalent values of $E_b/N_0$.  This makes $E_b/N_0$ the best SNR metric to use when comparing codes of different rate.
This paper further improves the $E_b/N_0$ performance by replacing the polar code with a tail-biting convolutional code (TBCC).  Specifically, a rate-1/5 TBCC is concatenated with a CRC optimized for the specific TBCC.  Lou \emph{et al.} \cite{Lou2015} introduced distance-spectrum-optimal (DSO) CRC's for zero-terminated convolutional codes.  Recently, Yang \emph{et al.} \cite{Yang2020}, \cite{Yang2021} presented an algorithm for finding DSO CRC's for tail-biting convolutional codes, which this paper employs to find the optimized CRCs used in this paper.

As is the case for polar codes,  TBCC/CRC performance is enhanced by optimizing the CRC length. Using DSO CRCs  and the CRC length that minimizes the FER for a specific $E_b/N_0$ yields a TBCC/CRC concatenated code that has better performance than the polar/CRC concatenation.  Decoder complexity and performance depend on list size, but the list decoder for the TBCC/CRC code required less run time on our computer for better performance than the list decoder for the polar/CRC concatenation.  We note that our optimal TBCC/CRC design has a higher rate of 32/215.  However, the use of repetition bits provides a code of rate 32/864 that has identical FER vs. $E_b/N_0$ performance to our original optimal TBCC/CRC design.

\subsection{Contributions}

This paper provides two ways to improve the FER  vs.  $E_b/N_0$ performance of the 5G PBCH. First, a shorter CRC can improve performance by allowing more frozen bits without sacrificing the needed error detection power.  Second, the concatenation of a polar code and CRC can be replaced by a TBCC concatenated with a CRC specifically optimized for the TBCC to provide better FER  vs.  $E_b/N_0$ performance with lower decoding complexity based on the run times of our simulations.

\subsection{Organization}
Section \ref{sec:Background} describes polar and TBCC codes.  Section \ref{sec:Optimal_CRC} presents results for the optimal CRC length for the TBCC and the trade-off between erasures and undetected errors.  Section \ref{sec:TBCC_Polar} compares the error rate and decoding complexities of the TBCC and polar codes and compares their performance to theoretical bounds.  Lastly, Section \ref{sec:Repetition} shows a near equivalence of our TBCC and a lower-rate TBCC attained by repeating bits to get the same rate as the polar code.

\section{Background}
\label{sec:Background}
This section describes the polar and TBCC codes that we will consider, presents a list decoder for TBCC/CRC and polar/CRC concatenated codes that employs parallel list decoding with exponentially increasing list sizes, and defines erasure failures and undetected errors. 

\subsection{Polar Codes}

Polar codes were first introduced by Arikan in \cite{ArikanPolar} as a code suitable to take advantage of the channel polarization paradigm that he discovered.  Polar codes compute the codeword by multiplying a message vector by a polar coding matrix.  The message vector contains both actual message bits and ``frozen'' bits that are set to a fixed value and do not convey information. The polarization paradigm ensures that the actual message bits have very high reliability while the frozen bits have a very low reliability, but are anyway set to a fixed value.  Polar codes have been shown to be able to achieve channel capacity for asymptotically long blocklengths; however, they are less reliable with short messages.

Also presented in \cite{ArikanPolar} is a proposed decoder for polar codes called a Successive Cancellation (SC) decoder.  This decoding algorithm decodes the received codeword one bit at a time, using previous decoded bits to help decide the current bit.  Frozen bits carry no information and are known to the decoder, so a decision only needs to be made on the message bits in the codeword.  As noted in \cite{SC_Performance}, the SC decoder is effective for decoding long messages, but is less effective for decoding the short messages used in 5G.  

This is addressed in 5G by using Successive Cancellation List (SCL) decoding in conjunction with a CRC \cite{3GPP38.212}.  Instead of making a hard decision on each message bit of the received codeword, the SCL algorithm \cite{ListSC} instead implements parallel decoders, one for each of a set of possible decisions about the previous bits.  When all the parallel decoders have each selected their distinct prospective codewords, these candidate codewords are then checked to see which pass the CRC check, and the most likely candidate that passes the check is selected.  If no candidate passes the CRC, then a decoding failure is reported.  

% Look into list size performance vs. complexity tradeoff for SCL decoding

The performance of SCL improves as the number of parallel decoders, i.e. the list size, is increased.  However, this improved performance comes with a significant complexity increase to support the parallel decoding \cite{ListSC}.  For a specified list size, SCL retains the most likely codeword candidates at each step.  In addition to the SCL algorithm, there have been many other proposed improvements to Arikan's initial SC decoder to increase decoding accuracy, decrease latency, and decrease complexity \cite{SimpleSC, ChenSCL, SC_Flip}.

The polar code used in this paper is the PBCH polar code from the 5G standard \cite{3GPP38.212}.  This code has 32 message bits and is encoded with a 24-bit CRC.  The 24-bit CRC has polynomial 0x1B2B117, with the most significant bit corresponding to the degree-24 term of the polynomial.  This 56-bit message and CRC is then encoded with a (512,56) polar code, and then the first 352 bits are repeated to arrive at a final 864-bit codeword, as illustrated in Fig. \ref{fig:5GPBCH}.

In addition to the 24-bit CRC that is used by the PBCH code, we also simulate this code with an 11-bit CRC provided in \cite{3GPP38.212} that is used for the physical uplink control channel (PUCCH) code, followed by a (512,43) polar code and the same bit repetition.  This 11-bit CRC has polynomial 0xE21.  Recently, Baicheva and Kazakov \cite{B&K}, \cite{B&K2} performed  an  analysis  on  the  5G  CRCs  and  presented alternative CRCs for polar codes than those in \cite{3GPP38.212}.  We also simulate the polar code with the 11-bit CRC 0xB5F and the 12-bit CRC 0x1395, provided by Peter Kazakov to achieve the best performance for this polar code.

A sequence of expected reliabilities is also provided in \cite{3GPP38.212} to select the bits having the lowest expected reliabilities to be frozen bits.   This paper uses SCL decoding for all polar codes we consider, as described in Section \ref{subsec:Decoder}.

\subsection{Tail-Biting Convolutional Codes}

In contrast to polar codes, convolutional codes have been in use for decades \cite{EliasCC}.  Convolutional codes can be used for transmitting streams of data with continuous decoding \cite{WeselEncyclopedia}, but they can also function as block codes.  One form of block convolutional codes is the class of TBCCs \cite{Ma1986}, which avoid the overhead incurred by zero termination.  

Our paper focuses on TBCCs because of their rate efficiency. The TBCC proposed in this paper as an alternative to polar coding for the 5G PBCH is taken from \cite{YangTBCC}.  This code is a rate-1/5 TBCC with 32 message bits, concatenated with a CRC.  The encoder has 8 memory elements with generator polynomials $(575, 623, 727, 561, 753)$ in octal. 

Lou \emph{et. al.} \cite{Lou2015} show how low undetected error rate performance of a convolutional code is dominated by the minimum distance spectrum of the code.  They present an algorithm to find DSO CRCs for a zero-terminated convolutional code by maximizing the minimum distance of the CC/CRC concatenated code.  This process is generalized to tail biting convolutional codes in \cite{Yang2020}.

In \cite{YangTBCC}, CRCs were designed for zero-terminated convolutional codes even though some simulations in \cite{YangTBCC} involved TBCCs.  In this paper, we deployed the algorithm described in \cite{Yang2020} to identify optimal CRCs for the TBCC implementation of $(575, 623, 727, 561, 753)$ with CRC lengths varying from 8 to 16 bits.  Table \ref{tab:CRC} provides the optimal CRC polynomials that resulted form our search.

\begin{table}[b]
    \caption{CRC polynomials for the TBCC at different CRC lengths.  Each polynomial is given in hexadecimal with the most significant bit corresponding to the highest order term.}

    \setlength\tabcolsep{2pt}    % vertical margin spacing
    \centering

    \begin{center}
    \begin{tabular}{|>{\centering}m{2cm}|>{\centering}m{0.5cm}|>{\centering}m{0.5cm}|>{\centering}m{0.5cm}|>{\centering}m{0.55cm}|>{\centering}m{0.6cm}|>{\centering}m{0.6cm}|>{\centering}m{0.68cm}|>{\centering}m{0.56cm}|m{0.73cm}|}
     \hline
     \vspace{0.09cm}
     CRC length & 8 & 9 & 10 & 11 & 12 & 13 & 14 & 15 & \hspace{0.1cm} 16 \\    % I cannot for the life of me center the last column in the above section, so I manually added whitespace instead
     \hline
     \vspace{0.09cm}
     CRC poly (hex) & 101 & 21F & 4D5 & A9D & 123B & 27C5 & 7CCF & 8441 & 18077 \\ 
     \hline
    \end{tabular}
    \end{center}

    \label{tab:CRC}
\end{table}

The Viterbi algorithm is a maximum-likelihood decoder for convolutional codes.  The decoder traverses the trellis identifying the most likely path to each state in the trellis based on the received codeword.  When multiple paths converge to the same state in the trellis, the decoder selects the most likely path, making an arbitrary choice to break ties.  At the end of the trellis, the decoder selects the most likely surviving path.  The Viterbi algorithm can be augmented to support parallel list decoding \cite{Seshadri1994}, where every state stores a list of the $L$ most likely paths instead of a single most likely path.

The TBCCs in this paper are decoded using an adaptive parallel list Viterbi algorithm (LVA) decoder based on the one described in \cite{Seshadri1994}.  The details of this decoder are also described in Section \ref{subsec:Decoder}.

\subsection{Parallel List Decoding with a Doubling List Size}
\label{subsec:Decoder}

This paper uses parallel list decoding with a doubling list size to explore the FER performance and decoding run time of both the 5G PBCH polar code and the proposed TBCC alternative.  The parallel list decoder is implemented as an SCL decoder for polar codes and using the LVA \cite{Seshadri1994} for convolutional codes. 

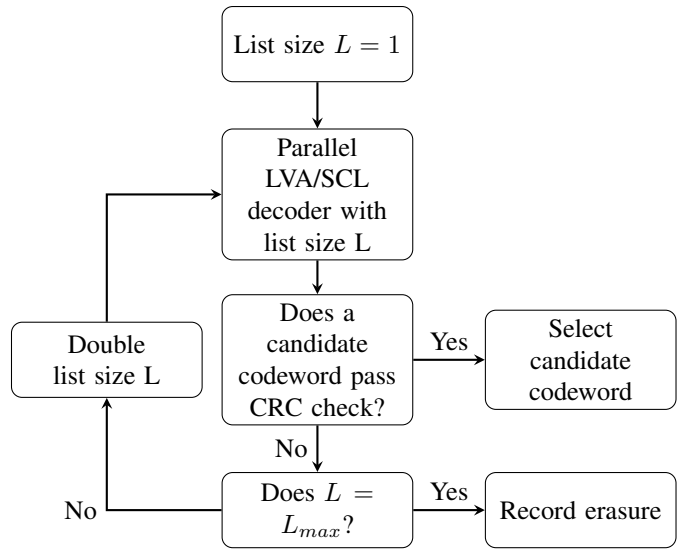
\begin{figure}

    \tikzstyle{box} = [rectangle, rounded corners, minimum width=2cm, minimum height=1cm,text centered, text width=2.3cm, draw=black]
    \tikzstyle{arrow} = [thick,->,>=stealth]

    \begin{tikzpicture}[node distance=2cm]
    
    \node (start) [box] {List size $L=1$};
    \node (decode) [box, below of=start] {Parallel LVA/SCL decoder with list size L};
    \node (CRC check) [box, below of=decode, yshift=-0.2cm] {Does a candidate codeword pass CRC check?};
    \node (success) [box, right of=CRC check, xshift=1.5cm] {Select candidate codeword};
    \node (list check) [box, below of=CRC check] {Does $L=L_{max}$?};
    \node (erasure) [box, right of=list check, xshift=1.5 cm] {Record erasure};
    \node (double L) [box, left of=CRC check, xshift=-0.8cm] {Double list size L};

    \draw [arrow] (start) -- (decode);
    \draw [arrow] (decode) -- (CRC check);
    \draw [arrow] (CRC check) -- node [anchor=south] {Yes} (success);
    \draw [arrow] (CRC check) -- node [anchor=east] {No} (list check);
    \draw [arrow] (list check) -- node [anchor=south] {Yes} (erasure);
    \draw [arrow] (list check) -| node [anchor=east] {No} (double L);
    \draw [arrow] (double L) |- (decode);
    
    \end{tikzpicture}

    \centering
    \caption{Block diagram of the adaptive parallel list decoder algorithm.  It starts with a list size of 1 and runs the parallel LVA or SCL algorithm.  This is repeated with the list size doubling every iteration until either a candidate codeword is found that passes the CRC check or the maximum list size is reached.}
    \label{fig:decoder_algorithm}
\end{figure}

This approach was proposed in \cite{AdaptiveSCL} for polar codes, where it was called an ``adaptive SCL'' decoder.  Each iteration of the algorithm acts like a parallel LVA or SCL decoder for the given list size.  If a message candidate is not found that passes the CRC check, then the list size doubles until either a codeword is found that passes the CRC check or the maximum list size is reached.  A block diagram of this list decoding algorithm is shown in Fig. \ref{fig:decoder_algorithm}.

There are two types of errors that can occur when using our list decoder.  An \emph{erasure} occurs when none of the decoded message candidates that the list decoder finds have a valid CRC when the decoder reaches the maximum list size.  An \emph{undetected error} occurs when one of the message candidates passes the CRC check, but it is not the same as the codeword sent.  In this paper, we use the sum of the erasure rate and the undetected error rate (UER) as a primary metric for performance, which we refer to as the Total Failure Rate (TFR).

\section{Optimal CRC Length for TBCC}
\label{sec:Optimal_CRC}

This section shows how a specified code has an optimal CRC length that minimizes the TFR. As an initial matter, a longer CRC should lead to an improved TFR at a fixed signal-to-noise ratio (SNR), but the longer CRC also reduces the rate.  To fairly compare the CRCs, we consider the TFR as a function of $E_b/N_0$, which accounts for the rate loss incurred by a longer CRC.

%At constant values of $E_b/N_0$, varying the length of the CRC in a TBCC/CRC concatenated codes, there exists a tradeoff between error detection and code-rate loss .  A longer CRC provides more protection for the codeword, decreasing the chance of an undetected error occurring.  However, increasing the CRC length also decreases the rate of the code. This decreases the channel bit SNR ($E_c/N_0$) at constant values of $E_b/N_0$, thus increasing the error probability.  

%In addition, it is expected that increasing the CRC length past a certain point has diminishing returns in terms of TFR.  With these combined effects, there is likely a CRC length for this TBCC/CRC code that minimizes TFR with respect to $E_b/N_0$.

\begin{figure}
    \centering
    \includegraphics[width=18pc]{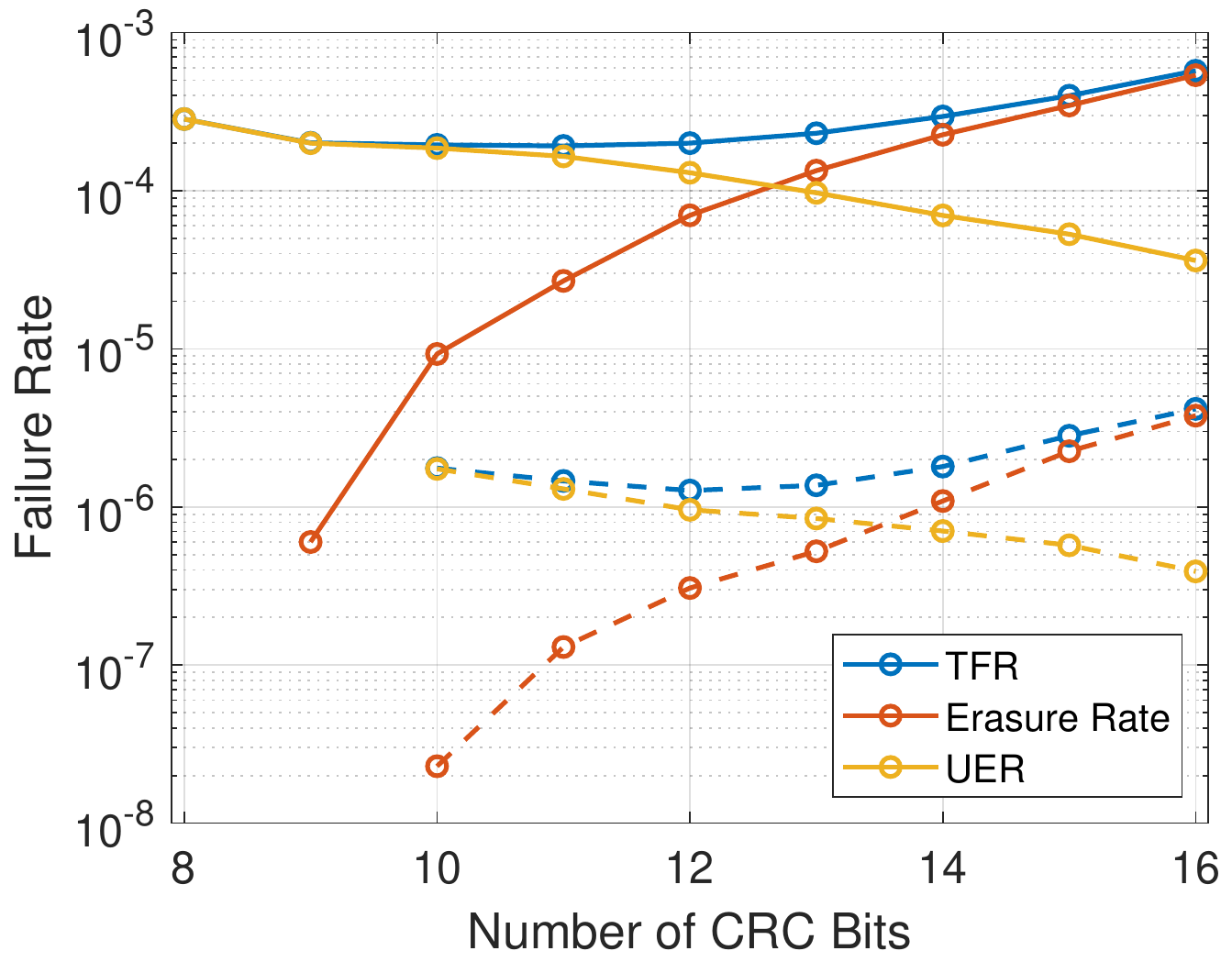}
    \caption{Plot of erasure failure rate, undetected error rate, and total failure rate vs. number of bits in the CRC for the TBCC.  The solid curves correspond to a $E_b/N_0$ of 2.5 dB, and the dashed curves have an $E_b/N_0$ of 3.5 dB.  A maximum list size of 2048 was used.  The CRC length that minimizes TFR is 11 bits at 2.5 dB and 12 bits at 3.5 dB, but nearby CRC lengths have nearly equivalent TFRs.}
    \label{fig:CRC_size}
\end{figure}

For two different fixed values of  $E_b/N_0$, 2.5 dB and 3.5 dB, Fig. \ref{fig:CRC_size} shows the erasure rate, UER, and TFR for simulating the TBCC/CRC with different CRC lengths with a maximum list size of 2048.  The CRCs used in these simulations are shown in Table \ref{tab:CRC}, and for each length the CRC used is optimal according to the procedure from \cite{Yang2020}.

Fig. \ref{fig:CRC_size} shows that, for this example, as CRC length increases, erasure rate monotonically increases and UER monotonically decreases.  Combining these two effects, TFR is convex or quasi-convex with a single global minimum.  

Thus, there is a CRC length that minimizes the TFR for a specified  value of $E_b/N_0$ and a specified maximum list size.  The optimal CRC length depends on the value of $E_b/N_0$.  At $E_b/N_0=2.5$ dB, the CRC length that minimizes the TFR is 11 bits.  At $E_b/N_0=3.5$ dB, the optimal length is 12 bits.  However, the difference in FER between 11-bit and 12-bit CRCs at these values of $E_b/N_0$ is almost negligible, and can change when the maximum list size is changed.

\section{Comparison of TBCCs and Polar Codes}
\label{sec:TBCC_Polar}

Consider the problem of transmitting a 32-bit message.  This section begins by comparing four polar code (PC) with CRC solutions to a rate-1/5 TBCC with an optimized CRC.  For the TBCC, the optimal CRCs for each length, shown in Table \ref{tab:CRC}, are designed according to \cite{Yang2020}.  The optimal CRC length, as discussed in Sec. \ref{sec:Optimal_CRC}, is also considered.

\subsection{TFR vs. $E_b/N_0$ and TFR vs. Run Time}
\begin{figure}
    \centering
    \includegraphics[width=18pc]{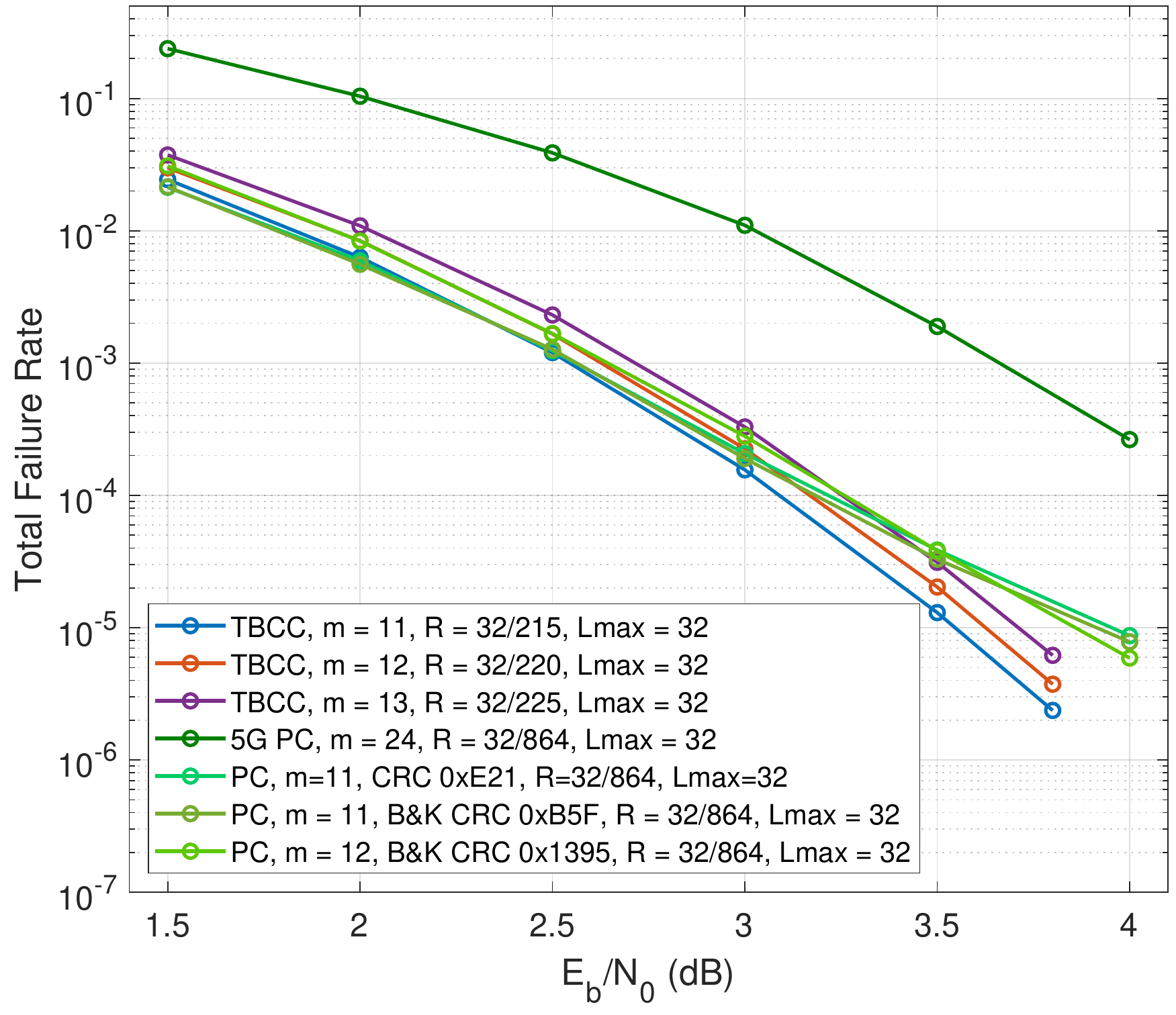}
    \caption{TFR vs. $E_b/N_0$ of TBCC and Polar Codes with various CRCs.  A maximum list size of 32 is used for all codes.  The TBCCs and $m=11$ and $m=12$ PCs achieve similar performance, with the PCs exhibiting a floor at high $E_b/N_0$.  These codes significantly outperform the $m=24$ 5G PC.}
    \label{fig:polar_TBCC}
\end{figure}

For a fixed maximum list size of 32, Fig. \ref{fig:polar_TBCC} shows TFR vs. $E_b/N_0$ for the 5G PC with 24-bit CRC solution, a rate-1/5 TBCC with CRCs of length $m=11, 12$, and $13$, and three additional PC-with-CRC solutions. The TBCC/CRC solutions all have similar performance, but the best performance is seen for CRC length $m=11$ for all $E_b/N_0$.  Note that the maximum list size $L_{max}=32$ is significantly  smaller than the maximum list size $L_{max}=2048$ considered in Sec. \ref{sec:Optimal_CRC} where the $m=12$ CRC is optimal at $E_b/N_0 = 3.5$ dB. 

In Fig. \ref{fig:polar_TBCC}, the 5G PC with the $m=24$ CRC specified in the 5G standard performs significantly worse than the the TBCC/CRC solutions.  To improve the PC performance, the CRC length was reduced to match the CRC length used for the convolutional code by using two different $m=11$ CRCs and one $m=12$ CRC.  The TFR vs. $E_b/N_0$ performance of the PC with the $m=11$ and $m=12$ CRCs is similar to that of the TBCC/CC solutions for this list size,  except for a TFR degradation seen at high $E_b/N_0$.

\begin{figure}
    \centering
    \includegraphics[width=19pc]{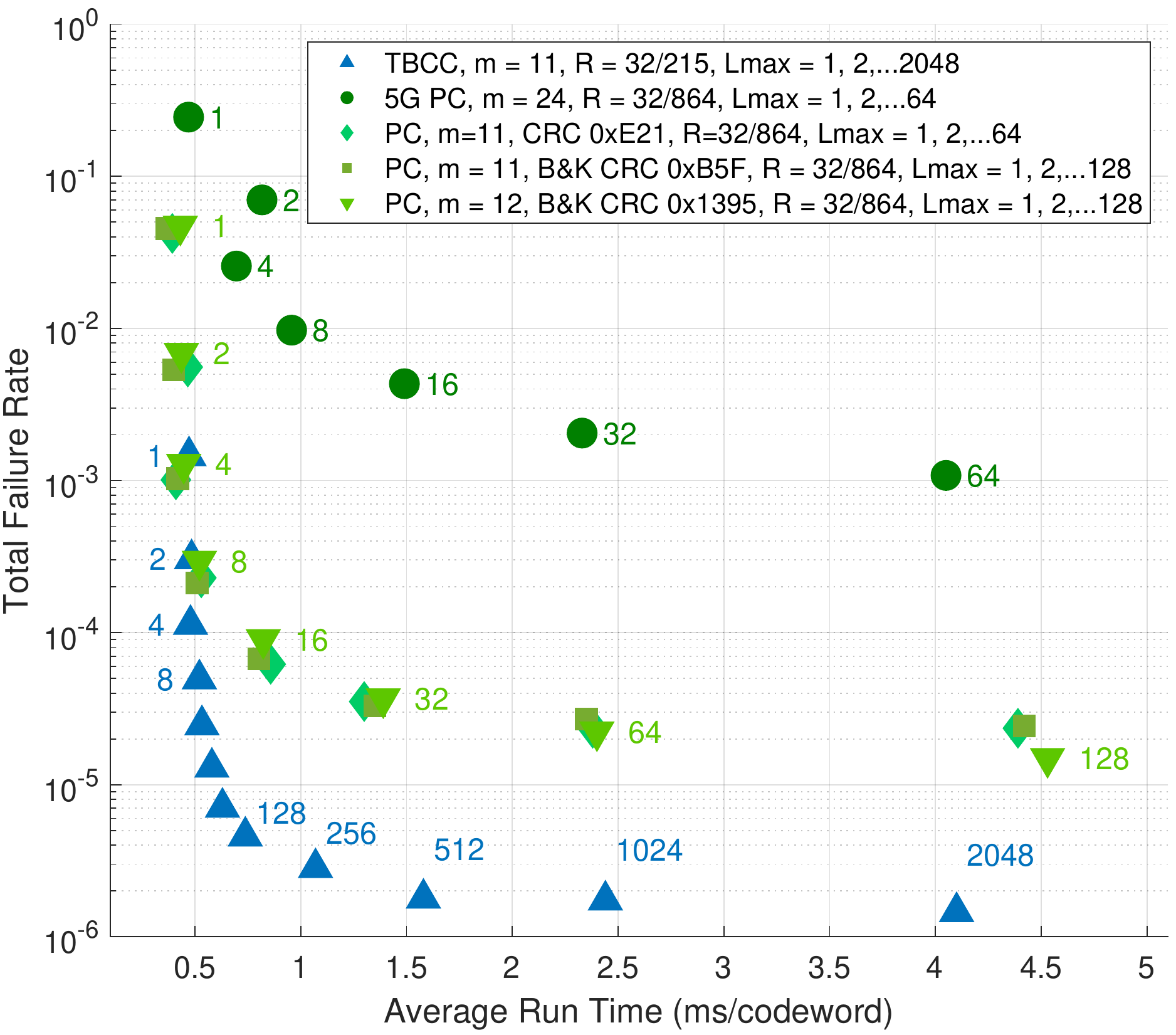}
    \caption{Average decoding run time in milliseconds of $m=11$ TBCC and all PCs vs. TFR at $E_b/N_0 = 3.5$dB.  Both $m=11$ PCs and the $m=12$ PC achieve far greater TFR performance at equivalent decoding run times than the 5G PC, and the $m=11$ TBCC performs even better.  There exists a trade-off between TFR and average decoding time when varying list size.  Eventually, increasing list size further does not provide any benefit to reducing TFR.}
    \label{fig:complexity}
\end{figure}

While all the curves in Fig. \ref{fig:polar_TBCC} used the same maximum list size, the decoders do not all have the same complexity or run time.  To explore complexity vs. TFR performance, Fig. \ref{fig:complexity} shows TFR  as a function of average simulation run time \footnote{All simulations were performed on a System76 Galaga Pro Ubuntu laptop with an Intel Core i7-8565U CPU @1.8GHz x 8 Processor.} at $E_b/N_0=3.5$dB for all the polar codes and the best-performing TBCC from Fig. \ref{fig:polar_TBCC}.  For both polar and TBCC codes, the decoders used C implementations of the exponentially increasing parallel list decoding paradigm of \cite{AdaptiveSCL}.  We tried to make both the TBCC and polar implementations as efficient as possible, but of course other implementations may result in different run-time comparisons.

%the polar list decoder We also compare the decoding complexity of TBCC and 5G polar codes. We measure decoding complexity by the average amount of time it took for the decoder to find a message candidate.  All complexity measurements were taken from the same computer in the same conditions, using a value of $E_b/N_0=3.5$dB.  Figure \ref{fig:complexity} shows the average decoding runtime vs. FER of the best performing TBCC, as well as the polar code with a few different CRCs.  We also vary the maximum list size to show how that affects decoding runtime and FER.
 
%The TBCC achieves a far lower FER are equivalent decoding runtimes compared to the polar code.  

In general, the TBCC is able to support a higher maximum list size and achieve a lower TFR for a specified run time.  For example, Fig. \ref{fig:complexity} shows that at an average decoding run time of 2.4 ms per decoded codeword, the TBCC achieves a TFR of $1.74 \times 10^{-6}$ using a maximum list size of 1024, while the best polar code with an 11-bit or 12-bit CRC only achieves  $2.23 \times 10^{-5}$ with a list size of 64.  At this run time, the 5G polar code with a 24-bit CRC achieves a TFR of $2.05 \times 10^{-3}$ with a list size of 32.

When the maximum list size is small in Fig. \ref{fig:complexity}, increasing the list size can dramatically reduce TFR while having a negligible impact on decoding run time. Further increases in list size provide diminishing returns in TFR performance but carry significant run time penalties.  Essentially, when the list size required to pass the CRC check is very large, the codeword that finally passes the CRC check will often be an undetected error failure so that TFR is not improved.

\begin{figure}
    \centering
    \includegraphics[width=19pc]{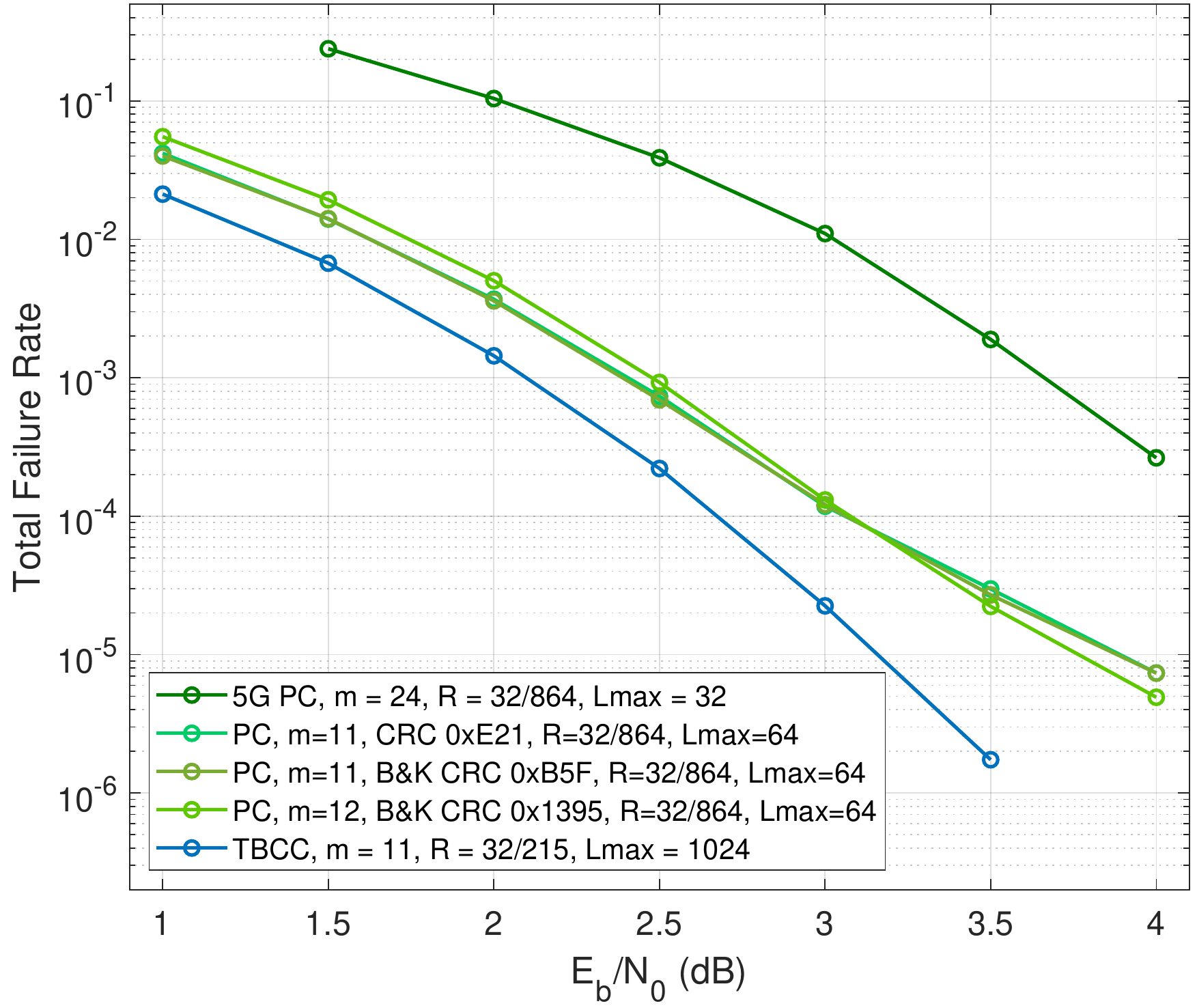}
    \caption{TFR vs. $E_b/N_0$ of $m=11$ TBCC, 5G PC, and all $m=11$ and $m=12$ PCs.  Each code is has a nearly equivalent average decoding runtime of around 2.4 ms per codeword, with the list sizes set according to Fig. \ref{fig:complexity} to achieve this.}
    \label{fig:equiv_dec}
\end{figure}

We did not consider list sizes that required more than 5 ms of average run time.  For the polar codes, this limited maximum list sizes to 64 or 128.  However, Fig. \ref{fig:complexity} shows that negligible improvement in TFR would be expected for larger maximum list sizes for these codes.

%Once the list size is large enough, increasing it further provides small benefit in TFR performance at the cost of a large decoding run time penalty.   

For the list sizes that resulted in run times of about\footnote{The run times are as follows: 2.44 ms per codeword for TBCC with $L_{max}=1024$, 2.38 ms per codeword for 5G 0xE21 $m=11$ PC with  $L_{max}=64$, 2.35 ms per codeword for B\&K 0xB5F $m=11$ PC, 2.4 ms per codeword for B\&K 0x1395 $m=12$ PC with $L_{max}=64$, and 2.45 ms per codeword for 5G $m=24$ PC with $L_{max}=32$}  2.4 ms at $E_b/N_0 = 3.5$dB, Fig. \ref{fig:equiv_dec} provides curves showing TFR vs. $E_b/N_0$ for the codes shown in Fig. \ref{fig:complexity}.

\begin{figure}
    \centering
    \includegraphics[width=19pc]{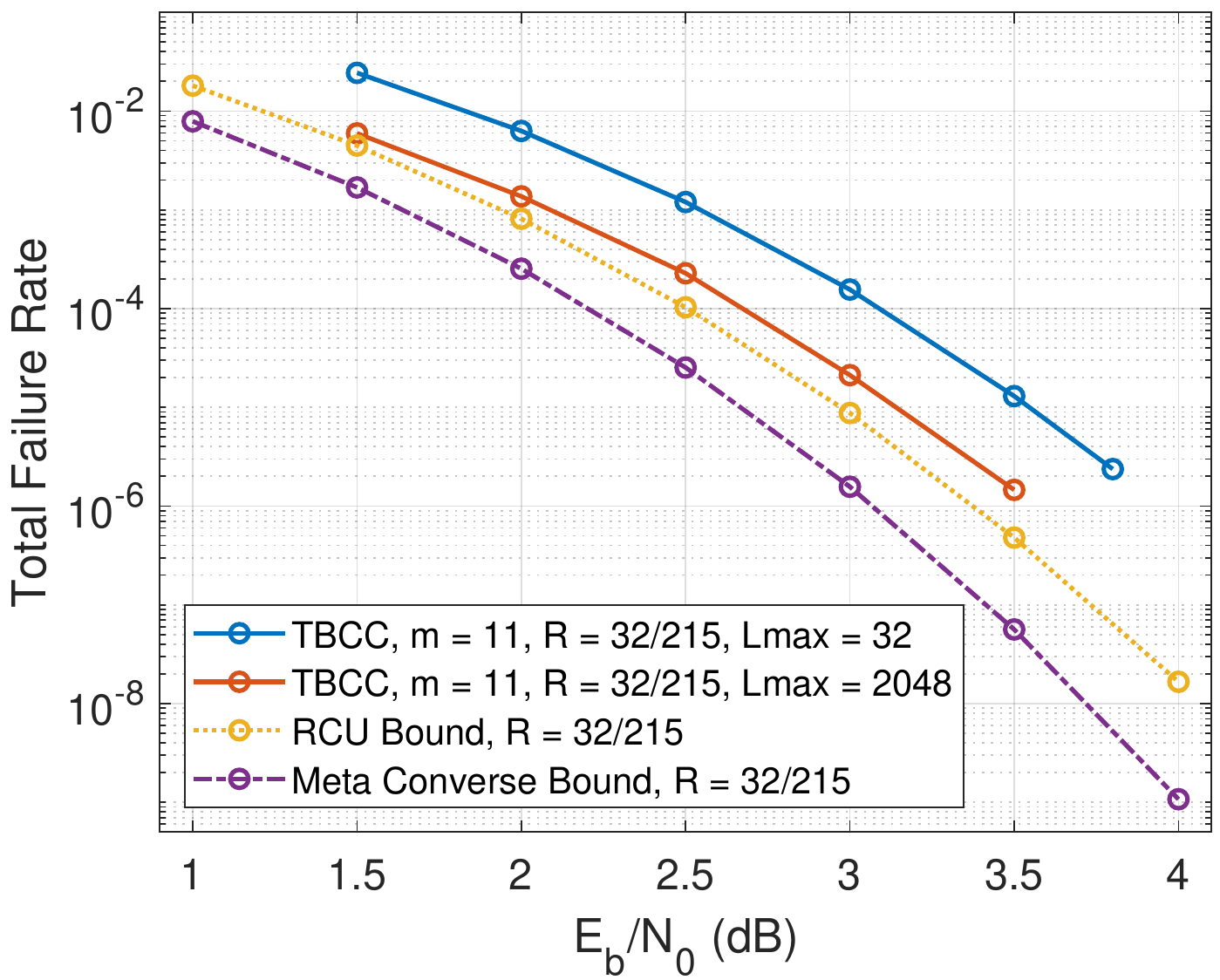}
    \caption{Plots of TFR, RCU Bound, and Meta Converse Bound of the $m=11$, Rate 32/215 TBCC vs. $E_b/N_0$ dB.  The $L_{max}=2048$ curve approaches very close to the RCU bound.  However, as shown in Fig. \ref{fig:complexity}, increasing the list size further is unlikely to improve TFR further.}
    \label{fig:m11}
\end{figure}

\begin{figure}
    \centering
    \includegraphics[width=19pc]{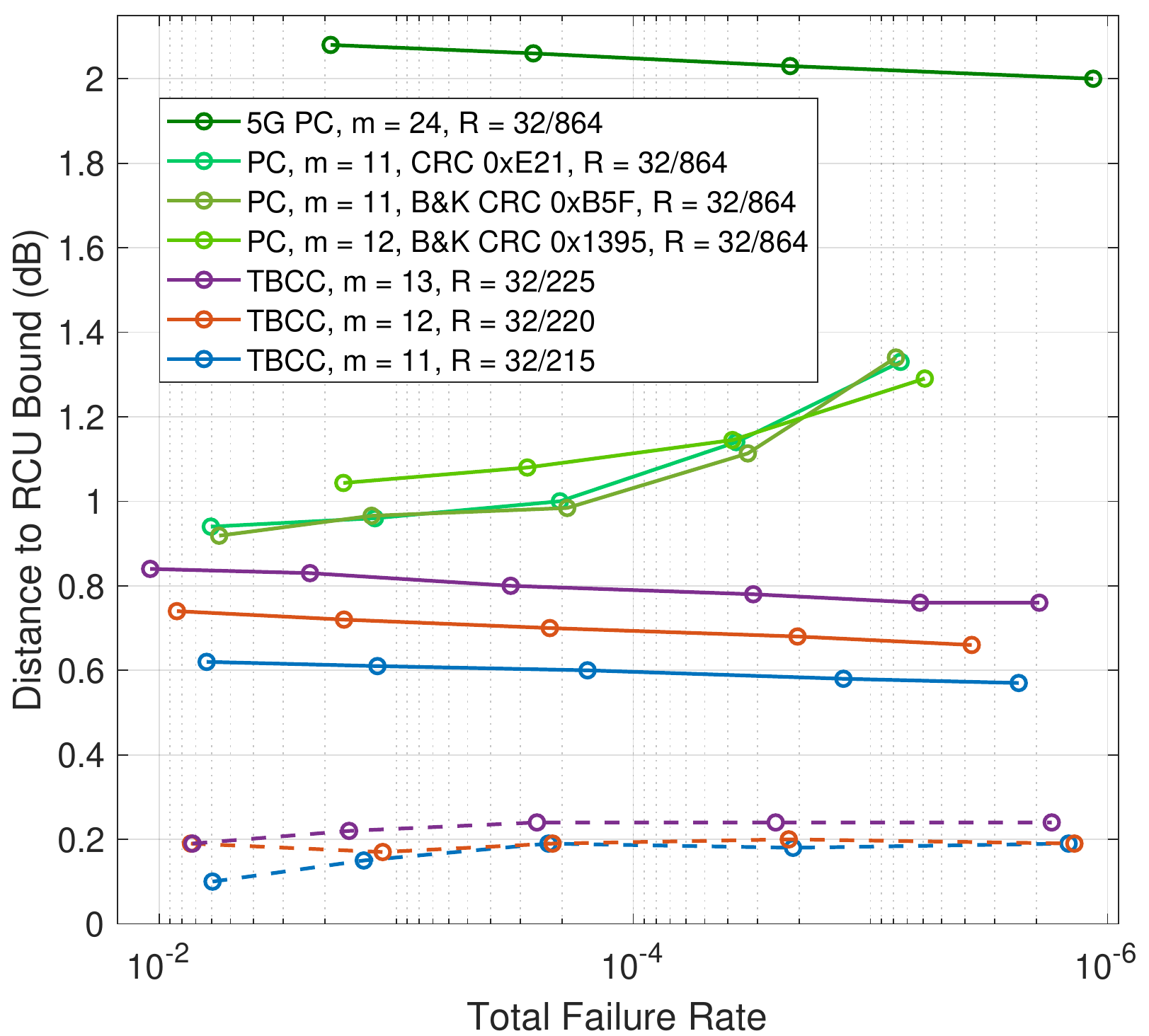}
    \caption{Gap to RCU bound vs TFR for all TBCC and polar codes simulated in this section.  Solid lines have $L_{max}=32$, and dashed lines have $L_{max}=2048$.   The TBCCs with $L_{max}=2048$ get very close to RCU Bound.  The TBCCs with $L_{max}=32$ also get closer to the RCU bound than all polar codes with $L_{max}=32$.  All $m=11$ and $m=12$ PCs outperform the $m=24$ 5G PC.}
    \label{fig:RCU}
\end{figure}

% Note: move Lmax info to caption

\subsection{Comparison to RCU and MC bounds}

The random coding union (RCU) bound \cite{Polyanskiy}, \cite{Saddlepoint}  can be used to provide an upper bound on the lowest achievable TFR for a code with a specified rate, blocklength, and $E_b/N_0$.  The meta-converse (MC) bound \cite{Polyanskiy}, \cite{Saddlepoint}  can be used to provide an lower bound on the lowest achievable TFR for a code with a specified rate, blocklength, and $E_b/N_0$.  Fig. \ref{fig:m11} shows TFR vs. $E_b/N_0$ for the TBCC with $m=11$, as well as the RCU bound and MC bound for this TBCC. At $L_{max}=2048$, the TBCC/CRC TFR approaches the RCU bound.

Fig. \ref{fig:RCU} shows the $E_b/N_0$ gap from the RCU bound for all TBCC and polar codes that were considered in the earlier figures. For TBCCs with $L_{max}=2048$, the $m=11$ CRC has the smallest gap across the entire TFR range. For example, at TFR of $1.46 \times 10^{-6}$ the gap to the RCU bound is 0.19 dB.

\section{ Exactly Matching Polar Rates via Repetition}
\label{sec:Repetition}

%%%%%%%%%%%%%%%%%%%%%%%%%
% Editing done up to here
%%%%%%%%%%%%%%%%%%%%%%%%%

The 5G Polar/CRC solution for the PBCH has 32/864, which is about 4 times lower than the TBCC/CRC solutions we propose.  However, the number of transmitted bits can be increased through repetition to 864 so that our TBCC/CRC solutions can be deployed with exactly the same rate as the PC/CRC solution of 5G.  Our proposed rate 1/5 TBCC with  an 11-bit CRC has an overall rate of 32/215. Repeating 211 of the 215 code bits four times, and repeating the remaining 4 bits five times times produces an overall rate of 32/864, which exactly matches the Polar/CRC solution.

A code resulting from  $M$ times repetition of every code bit has exactly the same TFR performance as the original code at a fixed value of $E_b/N_0$.  If we include the rate penalty but not the noise benefit of repeating those last four bits an extra time, we can bound the loss of the rate-32/864 to within 0.02 dB of the performance of the original rate-32/215 TBCC/CRC code or the equivalent  rate-32/860 code.  Our simulation of the rate-32/864 TBCC/CRC produced TFR vs. $E_b/N_0$ results that were indistinguishable from the simulation results for the rate-32/215 TBCC/CRC code.

\section{Conclusion}
\label{sec:Conclusion}

This paper shows two ways to improve the TFR vs. $E_b/N_0$ performance of the current 5G PBCH Polar/CRC code.  Reducing the CRC from 24 bits to 11 or 12 bits, which allows significantly more bits to be frozen, improves performance while still utilizing the paradigm of a polar code that uses CRC-aided list decoding.  Even better TFR vs. $E_b/N_0$ performance is achieved by replacing the polar code with a rate-1/5 TBCC with CRC-aided list decoding.  Repetition coding can be used to exactly match the rate of the current 5G PBCH code.

For the  TBCC,  the length of the CRC was optimized and the CRC polynomial was designed to optimize the distance spectrum of the concatenated code comprised of the TBCC and the CRC so as to minimize the TFR.  The TBCC/CRC solution has TFR vs. $E_b/N_0$ performance very close to the RCU bound when the maximum list size is allowed to be large, so that the TBCC/CRC solution is approaching the best performance that is theoretically guaranteed to be possible.  Notably, the decoding complexity for a given list size is lower for the TBCCs than for the polar codes, which allows the TBCC/CRC to benefit from larger maximum list sizes.

Our TBCC/CRC designs utilize \cite{Yang2020}, which shows how to optimize CRCs for use with TBCCs. We tried to find the best available CRC to use with the polar code. We considered the 11-bit CRC specified in the 5G standard \cite{3GPP38.212}, although that CRC is not specified for this polar code.  We also noted the work of Baicheva and Kazakov \cite{B&K}, \cite{B&K2} focused on designing CRCs to be used with polar codes and contacted them for assistance.  We are grateful to Peter Kazakov for providing the 11-bit and 12-bit CRCs that provided the best performance that we observed for CRC-aided decoding of the polar code.  These CRCs maximize the free distance of the CRC error detection code for the specific overall code lengths of 43 and 44 bits, respectively.

It remains an open problem to identify CRCs that optimize the distance spectrum of the concatenation of the polar and CRC code.  So, while this paper has identified best possible performance for CRC-aided decoding of the proposed TBCC, we expect the TFR vs. $E_b/N_0$ performance of CRC-aided decoding of the polar code can still be improved with a better CRC and may be able to more closely match that of the TBCC/CRC codes.  We are excited to find the best possible CRC for this polar code and report it in a future work.

\bibliography{paper_ref}
\end{document}